\documentclass[english]{article}
\usepackage[T1]{fontenc}
\usepackage[latin9]{inputenc}
\usepackage{verbatim}
\usepackage{float}
\usepackage{calc}
\usepackage{amsmath}
\usepackage{graphicx}
\usepackage{wasysym}
\usepackage{extarrows}
\usepackage{cite}

\usepackage{color}

\makeatletter
\usepackage{amssymb}
\usepackage{latexsym}
\usepackage{epsfig}
\usepackage{float}
\usepackage{dsfont}

\newcommand{\be}{\begin{equation}}
\newcommand{\ee}{\end{equation}}

\makeatother

\usepackage{babel}
\begin{document}
{}~ \hfill\vbox{\hbox{CTP-SCU/2018005}}\break
\vskip 3.0cm
\centerline{\Large \bf  Fixing three dimensional geometries from entanglement entropies of CFT$_2$}

\vspace*{10.0ex}
\centerline{\large Peng Wang, Houwen Wu and Haitang Yang  }
\vspace*{7.0ex}
\vspace*{4.0ex}
\centerline{\large \it Center for theoretical physics}
\centerline{\large \it Sichuan University}
\centerline{\large \it Chengdu, 610064, China} \vspace*{1.0ex}
\vspace*{4.0ex}

\centerline{pengw@scu.edu.cn, iverwu@scu.edu.cn, hyanga@scu.edu.cn}
\vspace*{10.0ex}
\centerline{\bf Abstract} \bigskip \smallskip
In this paper, we propose a method of fixing the leading behaviors of three dimensional geometries from the dual CFT$_2$ entanglement entropies. We employ only the holographic principle and do not use any assumption about the AdS/CFT correspondence and bulk geometry. Our strategy involves using both UV and IR-like CFT$_2$ entanglement entropies to fix the bulk geodesics. With a simple trick, the metric can be extracted from the geodesics. As examples, we fix the leading behaviors of the pure AdS$_3$ metric from the entanglement entropies of free CFT$_2$ and, more importantly, the BTZ black hole from the entanglement entropies of finite temperature CFT$_2$. Consequently, CFT$_2$ with finite size or topological defects can be determined through simple transformations. Following the same steps, in principle, the leading behaviors of  all three dimensional (topologically distinct)  holographic classical geometries   from the  dual CFT$_2$ entanglement entropies can be fixed.

\vfill
\eject
\baselineskip=16pt
\vspace*{10.0ex}
\tableofcontents

\section{Introduction}
 
As a manifestation of the non-local property of quantum mechanics,
quantum entanglement has attracted considerable interest in recent years.
The entanglement entropy (EE) measures the correlation between subsystems
and is one of the most distinct characteristics of quantum systems. Considering
the simplest configuration, a quantum system is divided into two subsystems:
$A$ and $B$. Thus, the total Hilbert space is accordingly decomposed into
$\mathcal{H}=\mathcal{H}_{A}\otimes\mathcal{H}_{B}$. Tracing out
the degrees of freedom of the region $B$, one obtains the reduced
density matrix of the region $A$: $\rho_{A}=\mathrm{Tr}_{\mathcal{H}_{B}}\rho$.
The EE of  region $A$ is evaluated using the von
Neumann entropy $S_{A}=-\mathrm{Tr}_{\mathcal{H}_{A}}\left(\rho_{A}\log\rho_{A}\right)$.
It is clear that $S_{A}=S_{B}$.

Motivated by the AdS/CFT correspondence \cite{Maldacena:1997re},
and the Bekenstein-Hawking entropy of black holes, Ryu and Takayanagi
(RT) proposed to identify the minimal surface area ending on the $d$
dimensional boundary of AdS$_{d+1}$ with the EE
of CFT$_{d}$ on the boundary of AdS$_{d+1}$ \cite{Ryu:2006bv, Ryu:2006ef,Hubeny:2007xt}.
In the case of $d=2$, the minimal surfaces are geodesics, and the
RT formula was verified extensively. For follow-ups and references,
please refer to a recent review \cite{Rangamani:2016dms}. The success
of the RT formula leads to an inspiring conjecture that gravity can
be interpreted as emergent structures, determined by the quantum entanglement
of the dual CFT \cite{VanRaamsdonk:2009ar,VanRaamsdonk:2010pw}. This
idea was further developed by Maldacena and Susskind to conjecture
an equivalence of Einstein-Rosen bridge (ER) and Einstein-Podolsky-Rosen
(EPR) experiment \cite{Maldacena:2013xja}, namely, ER=EPR. To justify
these conjectures, we must answer one question: Can the dual bulk
geometry, specifically the metric, be rebuilt from given CFT EEs?

Although, the dual geometries are conventionally believed to be asymptotic
AdS, no perfect method yet exists to fix the leading behaviors of the dual geometries
from the EEs of CFTs. One might believe fixing the leading behavior of the dual geometry from a   CFT$_{d}$ is trivial because they must possess the same $SO(2,d)$ symmetry.  This is not true as all
CFTs share the same symmetry and so do the dual geometries. The symmetry
argument is a necessary but not a sufficient condition to determine the dual
geometry of a CFT. However, the more important aspect is precisely the sufficient condition, i.e.
deriving  the dual geometries from CFTs. In other words, the simple symmetry argument
cannot carry beyond the vacuum configuration. Therefore, another systematic method applicable to all excited
states must be determined. Of course, because we aim to \emph{fix the leading behaviors of} the dual geometries,
we can not assume the geometries satisfying any dynamic equation, say, the
Einstein equation. The dynamic equation should also be derived from
the CFT information. The literature reveals several attempts, but
none of them can truly rebuild the dual geometries unambiguously. The
tensor network can only construct the discrete AdS \cite{Swingle:2009bg}.
Another major method utilizes integral geometry. The concept of
kinematic space is introduced, and the kinematic
space of AdS$_{3}$ is  argued to be dS$_{2}$, which can be read off from the Crofton
form defined as the second derivatives with respect to two different
points of the given EE of CFT$_{2}$ \cite{Czech:2015qta}.
However, proof has not been provided that AdS$_{3}$ is the only option to obtain dS$_{2}$
as the kinematic space. Moreover, this method applies only to the
static scenario naturally.

We can easily foresee that our journey to discover geometry reconstruction
is hindered by two difficulties:
\begin{itemize}
\item It is a standard homework to calculate the minimal surface from the
metric. Now that the CFT EE is identified with the
minimal surface, to obtain the geometry, we must extract
the metric from the minimal surfaces, which appears to be forbidden.
\item Reducing a higher dimensional theory to a
lower dimensional one is often not difficult after setting some limits or boundary conditions
to eliminate extra degrees of freedom. However, because the CFT$_{d}$
EE is identified with the minimal surface attached
on the boundary of the dual $d+1$ dimensional geometry,
to reconstruct the $d+1$ dimensional bulk geometry, we must determine a method
to uniquely fix the extra degrees of freedom, namely, the
bulk geodesic when $d=2$.
\end{itemize}
In a previous study \cite{Wang:2017bym}, we proposed an approach to
solving these two difficulties for $d=2$. A simple method exists 
to extract the metric from given geodesics, the minimal surfaces
for $d=2$. Let us consider a single geodesic $x=x\left(\tau\right)$
, $\tau\in\left[0,t\right]$ that connects points $x$ and $x'$
on manifold $M$, such that $x\left(0\right)=x$ and $x\left(t\right)=x^{\prime}$.
$L(x,x')$ is the length of the geodesic. Thus, the metric proves to
be

\begin{equation}
g_{ij}=-\:\underset{x\rightarrow x'}{\lim}\partial_{x^{i}}\partial_{x^{j^{\prime}}}\left[\frac{1}{2}L_{\mathrm{}}^{2}\left(x,x^{\prime}\right)\right].\label{eq:derive metric}
\end{equation}
For more details, a comprehensive review is provided in \cite{Poisson:2011nh}.
As an illustration, noting that along a geodesic, the norm of the
tangent vector $g_{ij}\dot{X}^{i}\dot{X}^{j}$ is constant; thus, for
very small distance $t\to0$, we have

\begin{equation}
\frac{1}{2}L^{2}(x,x')=\frac{1}{2}\left[\int_{0}^{t}d\tau\:\sqrt{g_{ij}\dot{X}^{i}\dot{X}^{j}}\right]^{2}\approx\frac{1}{2}\,\lim_{t\to0}\,g_{ij}\frac{\Delta x^{i}}{t}\frac{\Delta x^{j}}{t}t^{2}\approx\frac{1}{2}g_{ij}\Delta x^{i}\Delta x^{j}.
\end{equation}
Because quantity $\sigma\left(x,x^{\prime}\right)\equiv\frac{1}{2}L^{2}\left(x,x^{\prime}\right)$
is central to addressing the radiation back reaction of particles
moving in a curved spacetime, it has a specific name: Synge\textquoteright s
world function.

Therefore, what remains is to generalize the geodesics located on
the boundary to generic geodesics in the bulk. To fix the
expression of bulk geodesics, we observe that in addition to the typically
used UV EE, the IR-like EE of
the CFT is a prerequisite.

In the previous paper \cite{Wang:2017bym}, we addressed only the vacuum configuration,
i.e. the free CFT$_{2}$ with zero temperature and infinite length.
We showed explicitly the dual geometry must be AdS$_{3}$, as expected.
The purpose of this paper is to demonstrate that this approach also
works for excited states, specifically, the CFT$_{2}$ with finite
temperature, whose dual geometry is supposed to be the BTZ black hole.
It is well known that $3D$ gravity is topological as a consequence
of general relativity. Out of general geometries, the Einstein
equation selects those with no local degrees of freedom to describe
gravity. However, because we aim to fix the leading behavior of the dual geometry, we can
not use any results from general relativity. Therefore, the local agreement
of the dual geometries should be unknown and revealed
by the derived metrics from the Es. Because the EE is a non-local quantity, we
cannot directly transform the EE of the free
CFT to that of the finite temperature CFT as they have different
topologies. In contrast, when we fix the leading behaviors of the BTZ black hole from the
finite temperature CFT, we can easily extend the result to the finite
size CFT under the transformation $\beta=-iL$ or CFT with topological
defects under transformation $\beta=-iL/\gamma_{\mathrm{con}}$, because
they have the same topology: a cylinder. More importantly, the BTZ
derivation indicates that, with our approach, the leading behaviors of
all $3D$ classical (topologically distinct) geometries from the EEs of the dual CFT$_{2}$ can be fixed.

The reminder of this paper is outlined as follows. In Sec. 2, we
briefly review some useful results of CFT EEs,
which aids us in determining the geodesic length in the bulk geometry.
In Sec. 3, we show how to fix the leading behavior of
the BTZ spacetime from the EEs of the finite temperature
CFT. In Sec. 4, we present some inspirations and conjectures.

\section{Some useful results of CFT EE}

In this section, based on refs. \cite{Calabrese:2004eu,Calabrese:2009qy},
we briefly summarize some results of CFT$_{2}$ EEs that we will use in the remainder of the paper. For a quantum system
consisting of two parts, $A$ and $B$, the EE of subsystem $A$ is defined by the Von Neumann entropy:

\begin{equation}
S_{EE}=-\mathrm{Tr}_{\mathcal{H}_{A}}\left(\rho_{A}\log\rho_{A}\right),
\end{equation}

\noindent where reduced density matrix $\rho_{A}=\mathrm{Tr}_{\mathcal{H}_{B}}\rho$
and $\rho=\left|\Psi\right\rangle \left\langle \Psi\right|$. In QFT, calculating the Von Neumann entropy directly  is often difficult.
Alternatively, we use the ``replica trick'' to calculate $\mathrm{Tr}\rho_{A}^{n}$;
thus,

\begin{equation}
S_{EE}=-\underset{n\rightarrow1}{\lim}\frac{\partial}{\partial n}\mathrm{log\:Tr}\rho_{A}^{n}.\label{eq:density to S}
\end{equation}

\noindent Considering a 1+1 dimensional Euclidean QFT with a local
field $\phi\left(t_{E},x\right)$, we obtain

\begin{equation}
\mathrm{Tr}\rho_{A}^{n}=\frac{1}{\left(Z_{1}\right)^{n}}\int_{\left(t_{E},x\right)\in\mathcal{R}_{n}}\mathcal{D}\phi e^{-S_{E}\left(\phi\right)}\equiv\frac{Z_{n}\left(A\right)}{\left(Z_{1}\right)^{n}},
\end{equation}

\noindent where $Z_{n}\left(A\right)$ is the partition function on
$n$-sheeted Riemann surface $\mathcal{R}_{n}$, and $Z_{1}$ is the
vacuum partition function on $\mathcal{R}_{2}$. The partition function
is given by the two-point function of twist operators $\mathcal{T}$
and $\tilde{\mathcal{T}}$. For an infinitely long system, when fixing
$t_{E}=it=0$,

\begin{equation}
Z_{n}\left(A\right)=\left\langle \mathcal{T}_{n}\left(u,0\right)\tilde{\mathcal{T}}_{n}\left(v,0\right)\right\rangle _{\mathbb{C}}=\frac{1}{\left|u-v\right|^{2\Delta}},\label{eq:twist}
\end{equation}

\noindent where $\triangle=\frac{c}{12}\left(n-\frac{1}{n}\right)$
is the scaling dimension and $c$ is the central charge. Therefore, the EE is

\begin{eqnarray}
S_{EE} & = & -\underset{n\rightarrow1}{\lim}\frac{\partial}{\partial n}\mathrm{\log\:Tr}\rho_{A}^{n}=\frac{c}{3}\log\frac{\triangle x}{a}+c_{1}^{\prime},\label{eq:EE infinite}
\end{eqnarray}

\noindent where $c_{n}^{\prime}\equiv\log c_{n}/\left(1-n\right)$,
and $a$ is an energy cut-off that ensures the factor inside the
$log$ is dimensionless. $\Delta x=x'-x$ is the size of entangling
region $A$.

We can easily develop this result to other geometric background by
utilizing conformal transformations $z^{\prime}\rightarrow z=z\left(z^{\prime}\right)$
on two-point functions:

\begin{equation}
\left\langle \mathcal{T}_{n}\left(z_{1}^{\prime},\bar{z}_{1}^{\prime}\right)\tilde{\mathcal{T}}_{n}\left(z_{2}^{\prime},\bar{z}_{2}^{\prime}\right)\right\rangle =\left(\frac{\partial z_{1}}{\partial z_{1}^{\prime}}\frac{\partial z_{2}}{\partial z_{2}^{\prime}}\right)^{\triangle}\left(\frac{\partial\bar{z}_{1}}{\partial\bar{z}_{1}^{\prime}}\frac{\partial\bar{z}_{2}}{\partial\bar{z}_{2}^{\prime}}\right)^{\triangle}\left\langle \mathcal{T}_{n}\left(z_{1},\bar{z}_{1}\right)\tilde{\mathcal{T}}_{n}\left(z_{2},\bar{z}_{2}\right)\right\rangle .\label{eq:CFT trans}
\end{equation}

\noindent For example, to calculate the EE of a
CFT$_{2}$ at finite temperature $2\pi\beta^{-1}$, we map infinitely
long cylinder $z^{\prime}$ to plane $z\left(z^{\prime}\right)$
using the following transformation:

\begin{equation}
z^{\prime}\rightarrow z\left(z^{\prime}\right)=e^{\frac{2\pi z^{\prime}}{\beta}}.\label{eq:finite trans}
\end{equation}

\noindent The two-point function of the finite temperature CFT$_{2}$
is

\begin{eqnarray}
\left\langle \mathcal{T}_{n}\left(u^{\prime},0\right)\tilde{\mathcal{T}}_{n}\left(v^{\prime},0\right)\right\rangle _{\mathbb{C}} & = & \left(\frac{\partial}{\partial u^{\prime}}e^{\frac{2\pi u^{\prime}}{\beta}}\right)^{\triangle}\left(\frac{\partial}{\partial v^{\prime}}e^{\frac{2\pi v^{\prime}}{\beta}}\right)^{\triangle}\left\langle \mathcal{T}_{n}\left(u,0\right)\tilde{\mathcal{T}}_{n}\left(v,0\right)\right\rangle _{\mathbb{C}}\nonumber \\
 & = & \left|\frac{\beta}{\pi}\sinh\frac{\pi\triangle x}{\beta}\right|^{-\frac{c}{6}\left(n-\frac{1}{n}\right)}.
\end{eqnarray}

\noindent Therefore, the EE is given by

\begin{equation}
S_{EE}=\frac{c}{3}\log\left(\frac{\beta}{\pi a}\sinh\frac{\pi\triangle x}{\beta}\right)+c_{1}^{\prime}.
\end{equation}

\noindent Note that although the two-point function
of the finite temperature CFT$_{2}$ can be obtained from that of
the free CFT$_{2}$ using a conformal map, no coordinate transformation
exists to connect their EEs. This is because EE
is a global quantity associated with the topology, whereas these
two systems clearly have different topologies. In contrast, because
a finite size system has the same topology as a finite temperature
system, the EE of a finite size system is obtained
by replacing $\beta\rightarrow L_{S}$ and imposing the periodic boundary
condition

\begin{equation}
S_{EE}=\frac{c}{3}\log\left(\frac{L_{S}}{\pi a}\sin\left(\frac{\pi\triangle x}{L_{S}}\right)\right)+c_{1}^{\prime},
\end{equation}

\noindent where $L_{S}$ is the circumference of the given system.

When deriving the BTZ geometry, we also require the EE of the boundary conformal field theory (BCFT).  The BCFT
is a CFT whose boundary satisfies conformally invariant boundary conditions.
Considering an one dimensional semi-infinite long system $x\in\left[0,\infty\right)$,
the boundary is clearly located at $x=0$. The $n$-sheeted Riemann
surface $\mathcal{R}_{n}$ now consists of $n$ copies in the region
of $x\geq0$. The transformation from complex coordinates on $\mathcal{R}_{n}$
to $\mathbb{C}$ is $w\rightarrow z\left(w\right)=\left(\left(w-il\right)/\left(w+il\right)\right)^{1/n}$.
The partition function $Z_{n}\left(A\right)$ on the $n$-sheeted
Riemann surface $\mathcal{R}_{n}$ becomes the one-point function
of twist operator $\mathcal{T}$ . For any primary operator $\mathcal{O}$,
the one-point function is

\begin{equation}
\left\langle \mathcal{O}\left(z\right)\right\rangle =\frac{1}{\left(2\mathrm{Im}\:z\right)^{\triangle}}.
\end{equation}

\noindent The scaling dimension of $\mathcal{T}$ still equals $\triangle=\frac{c}{12}\left(n-\frac{1}{n}\right)$.
Therefore, we obtain

\begin{equation}
\left\langle \mathcal{\mathcal{T}}\left(il\right)\right\rangle =\frac{1}{\left(2l\right)^{\frac{c}{12}\left(n-\frac{1}{n}\right)}}.\label{eq:one-point of BCFT}
\end{equation}

\noindent Thus, we straightforwardly observe that

\begin{eqnarray*}
S_{EE} & = & -\underset{n\rightarrow1}{\lim}\frac{\partial}{\partial n}\mathrm{\log\:Tr}\rho_{A}^{n}=\frac{c}{6}\log\frac{2\triangle x}{a}+\tilde{c}_{1}^{\prime}.
\end{eqnarray*}

\noindent Applying the transformations (\ref{eq:finite trans}) onto
the one-point function (\ref{eq:one-point of BCFT}), we obtain

\begin{equation}
\left\langle \mathcal{\mathcal{T}}\left(il^{\prime}\right)\right\rangle =\left|\frac{\beta}{\pi}\sinh\frac{2\pi\triangle x}{\beta}\right|^{-\frac{c}{12}\left(n-\frac{1}{n}\right)}.
\end{equation}

\noindent Therefore, the EE of the BCFT at finite temperature
is

\begin{equation}
S_{EE}=\frac{c}{6}\log\left(\frac{\beta}{\pi a}\sinh\frac{2\pi\triangle x}{\beta}\right)+\tilde{c}_{1}^{\prime}.\label{eq:BCFT}
\end{equation}

\noindent Note that $\Delta x$ here is the entanglement length of
the BCFT, which is half the entanglement length of the corresponding
CFT.

Often, when one mentions the EE,
he/she really refers to the UV EE, which is precisely
what we have discussed thus far. However, when a free CFT is perturbed
by a relevant operator, the correlation length $\xi$ (IR cut-off)
takes a finite value. In the IR region $\Delta x\gg\xi$, the UV EE (\ref{eq:EE infinite}) is no longer valid, and an IR EE exists. The simplest method to calculate the IR EE is to consider the action

\begin{equation}
S=\int d^{2}x\left(\frac{1}{2}\left(\partial_{\mu}\varphi\right)^{2}+\frac{1}{2}m^{2}\varphi^{2}\right),
\end{equation}

\noindent where $m\rightarrow0$. Partition function $Z_{n}\left(A\right)$
on $n$-sheeted Riemann surface $\mathcal{R}_{n}$ can be calculated
with the identity \cite{Calabrese:2009qy}

\begin{equation}
\frac{\partial}{\partial m^{2}}\log Z_{n}\left(A\right)=-\frac{1}{2}\int d^{2}xG_{n}\left(\mathbf{x},\mathbf{x}\right),
\end{equation}

\noindent where $G_{n}\left(\mathbf{x},\mathbf{x}\right)$ is the
two-point function on $\mathcal{R}_{n}$, satisfying the equation
of motion $\left(-\nabla^{2}+m^{2}\right)G_{n}\left(\mathbf{x},\mathbf{x}^{\prime}\right)=\delta^{2}\left(\mathbf{x}-\mathbf{x}^{\prime}\right)$.
Thus,

\begin{equation}
\frac{\partial}{\partial m^{2}}\log\frac{Z_{n}\left(A\right)}{\left(Z_{1}\right)^{n}}=\frac{1}{24m^{2}}\left(n-\frac{1}{n}\right).
\end{equation}

\noindent Integrating $m^{2}$ on both sides, we obtain

\begin{equation}
\log\frac{Z_{n}\left(A\right)}{\left(Z_{1}\right)^{n}}=\frac{\log a^{2}m^{2}}{24}\left(n-\frac{1}{n}\right)\rightarrow\frac{Z_{n}\left(A\right)}{\left(Z_{1}\right)^{n}}=\left(ma\right)^{\frac{1}{12}\left(n-\frac{1}{n}\right)}.
\end{equation}

\noindent Therefore, the IR EE is

\begin{eqnarray}
S_{EE}^{IR} & = & -\underset{n\rightarrow1}{\lim}\frac{\partial}{\partial n}\log\:\frac{Z_{n}\left(A\right)}{\left(Z_{1}\right)^{n}}\nonumber \\
 & = & -\underset{n\rightarrow1}{\lim}\frac{\partial}{\partial n}\left[\left(ma\right)^{\frac{1}{12}\left(n-\frac{1}{n}\right)}\right]\nonumber \\
 & = & \frac{1}{6}\log\frac{\xi}{a},\label{eq:EE IR}
\end{eqnarray}

\noindent where $c=1$ for one field $\varphi$, and we introduce
IR cut-off $\xi=m^{-1}$.

The time dependent EEs can be calculated by completely
the same procedure. At each step, we simply include the time-like
variable to obtain

\begin{eqnarray}
\mathrm{infinite\quad system:} &  & S_{EE}\left(t\right)=\frac{c}{3}\log\frac{\sqrt{\left(\triangle x\right)^{2}-\left(\triangle t\right)^{2}}}{a},\label{eq:time uv}\\
\mathrm{finite\quad temperature:} &  & S_{EE}\left(t\right)=\frac{c}{3}\log\left(\frac{\beta}{2\pi a}\left[\sqrt{2\cosh\left(\frac{2\pi\triangle x}{\beta}\right)-2\cosh\left(\frac{2\pi\triangle t}{\beta}\right)}\right]\right),
\label{eq: FT}
\end{eqnarray}

\noindent which are well-defined when two points are space-like separated.

\section{BTZ spacetime from entanglement}

We now consider a finite temperature CFT$_{2}$. Two energy scales exist: UV cut-off $a$ and temperature $T^{-1}=\frac{\beta}{2\pi}\equiv\beta_{H}$.
We use notation $\beta_{H}\equiv\frac{\beta}{2\pi}$ for simplicity
in remainder of the paper. The temperature introduces a natural upper
bound for the energy generated extra dimension: $y\le\beta_{H}$.

Our first step is to determine the most general expression of the bulk geodesic of the dual geometry for the finite temperature CFT$_{2}$ and then fix the arbitrary functions using known CFT data. Two immediate restrictions occur:

\begin{enumerate}
\item Since we are fixing the dual geometry of finite temperature CFT$_2$, when ending on the boundary, the geodesic length must match the EE of the finite temperature CFT$_{2}$  given by (\ref{eq: FT})
\begin{eqnarray}
\frac{L_{\mathrm{boundary}}}{R} & = & \log\left(\frac{\beta_{H}^{2}}{a^{2}}\left[2\cosh\left(\frac{\triangle x}{\beta_{H}}\right)-2\cosh\left(\frac{\triangle t}{\beta_{H}}\right)\right]\right),\label{eq:BTZ boundary L}
\end{eqnarray}

\item As $\beta_{H}\rightarrow\infty$ ($T\to0$), the finite temperature
CFT$_{2}$ reduces to the free CFT$_{2}$. In the previous work,
we have shown that the dual geometry of free CFT$_{2}$ 
is  pure AdS$_{3}$ \cite{Wang:2017bym}. Therefore, the dual geometry of finite temperature CFT$_2$ must have pure AdS$_3$ as its $\beta_H\to \infty$ limit. In other words, when $\beta_{H}\rightarrow\infty$ ($T\to0$),
the dual geometry geodesic of  finite temperature
CFT$_{2}$ must be
\begin{equation}
\cosh\left(\frac{L_{\mathrm{bulk}}}{R}\right)=\frac{\left(\triangle x\right)^{2}-\left(\triangle t\right)^{2}+y^{2}+y^{\prime2}}{2yy^{\prime}}.\label{eq:pure AdS L}
\end{equation}
\end{enumerate}

\noindent Moreover, based on the holographic principle, the energy cut-off generates
extra dimension $a\to y$. As these requirements,
the most general expression of the bulk geodesic of the dual geometry
for the finite temperature CFT$_{2}$ can only take the form

\begin{equation}
\cosh\left(\frac{L_{\mathrm{bulk}}}{R}\right)=\frac{\beta_{H}^{2}}{yy^{\prime}}\left[f\left(x,x';y,y^{\prime};t,t'\right)\cosh\left(\frac{\triangle x}{\beta_{H}}\right)-g\left(x,x';y,y^{\prime};t,t'\right)\cosh\left(\frac{\triangle t}{\beta_{H}}\right)\right].\label{eq:BTZ cosh}
\end{equation}

\noindent where $f\left(x,x';y,y^{\prime};t,t'\right)$ and $g\left(x,x';y,y^{\prime};t,t'\right)$
are the regular functions to be determined, and they must be invariant
under $\left(x^{\prime},y^{\prime},t^{\prime}\right)\leftrightarrow\left(x,y,t\right)$.
The $cosh$ on the LHS of eqn. (\ref{eq:BTZ cosh}) is determined
using eqn. (\ref{eq:pure AdS L}), and the function form on the RHS of
eqn. (\ref{eq:BTZ cosh}) is determined using eqn. (\ref{eq:BTZ boundary L}).
We do not place the term of $\cosh\left(\frac{\triangle y}{\beta_{H}}\right)$
in eqn. (\ref{eq:BTZ cosh}) because its existence, if any, can be
absorbed in undetermined functions $f\left(x,x';y,y^{\prime};t,t'\right)$
and $g\left(x,x';y,y^{\prime};t,t'\right)$. Similarly, factor
$\frac{1}{yy'}$ outside the bracket is simply for convenience. Therefore,
the aim is the same as the free CFT scenario: we apply various constraints
to determine functions $f$ and $g$ and then use $L_{\mathrm{bulk}}$
to obtain the metric.

We stress again here why we assume some arbitrary metric. Our aim is to fix the dual geometry of CFT, which belongs to kinematics. The next much harder and  more important step is to derive the dynamic equation, i.e., Einstein equation, from CFT data. As gravity is the geometry respecting  the Einstein equation, we can safely claim that gravity emerges from quantum entanglement.

\vspace{2em}

\noindent \textbf{Step 1:} When $\beta_{H}\gg y=y^{\prime}=a$, $L_{\mathrm{bulk}}$
must reduce to $L_{\mathrm{boundary}}$, given by eqn. (\ref{eq:BTZ boundary L}),

\begin{eqnarray}
L_{\mathrm{bulk}} & = & R\log\left(\frac{\beta_{H}^{2}}{yy^{\prime}}\left[f\left(x,x';y,y^{\prime};t,t'\right)2\cosh\left(\frac{\triangle x}{\beta_{H}}\right)-g\left(x,x';y,y^{\prime};t,t'\right)2\cosh\left(\frac{\triangle t}{\beta_{H}}\right)\right]\right)\nonumber \\
 & \rightarrow & R\log\left(\frac{\beta_{H}^{2}}{a^{2}}\left[2\cosh\left(\frac{\triangle x}{\beta_{H}}\right)-2\cosh\left(\frac{\triangle t}{\beta_{H}}\right)\right]\right).
\end{eqnarray}

\noindent Therefore, as $\beta_{H}\gg y$ and $y^{\prime}$, we obtain

\begin{eqnarray}
f\left(x,x';y,y^{\prime};t,t'\right) & = & 1+\mu_{1}\left(x,x';t,t'\right)\left(\frac{y}{\beta_{H}}+\frac{y^{\prime}}{\beta_{H}}\right)+\mu_{2}\left(x,x';t,t'\right)\left(\frac{y}{\beta_{H}}+\frac{y^{\prime}}{\beta_{H}}\right)^{2}+\ldots\nonumber \\
 &  & +\rho_{1}\left(x,x';t,t'\right)\left(\frac{yy^{\prime}}{\beta_{H}^{2}}\right)+\rho_{2}\left(x,x';t,t'\right)\left(\frac{yy^{\prime}}{\beta_{H}^{2}}\right)^{2}+\ldots,\nonumber \\
\nonumber \\
g\left(x,x';y,y^{\prime};t,t'\right) & = & 1+\bar{\mu}_{1}\left(x,x';t,t'\right)\left(\frac{y}{\beta_{H}}+\frac{y^{\prime}}{\beta_{H}}\right)+\bar{\mu}_{2}\left(x,x';t,t'\right)\left(\frac{y}{\beta_{H}}+\frac{y^{\prime}}{\beta_{H}}\right)^{2}+\ldots\nonumber \\
 &  & +\bar{\rho}_{1}\left(x,x';t,t'\right)\left(\frac{yy^{\prime}}{\beta_{H}^{2}}\right)+\bar{\rho}_{2}\left(x,x';t,t'\right)\left(\frac{yy^{\prime}}{\beta_{H}^{2}}\right)^{2}+\ldots,
\end{eqnarray}

\noindent where $\mu_{i}$, $\rho_{i}$, $\nu_{i}$, $\bar{\mu}_{i}$,
$\bar{\rho}_{i}$, and $\bar{\nu}_{i}$ are the regular and bounded functions
regardless of the values of $\triangle x$ and $\triangle t$.

\vspace{2em}

\noindent \textbf{Step 2:} As $\beta_{H}\rightarrow\infty$ or $\beta_{H}\gg\triangle x$,
$\triangle t$, $y$ and $y^{\prime}$, the general expression (\ref{eq:BTZ cosh})
must match the pure AdS$_{3}$ background (\ref{eq:pure AdS L}).
From step 1, we know the leading term of $f$ and $g$ is the unit.
Therefore, we obtain

\begin{eqnarray}
 &  & \cosh\left(\frac{L_{\mathrm{bulk}}}{R}\right)\nonumber \\
 & \simeq & \frac{\beta_{H}^{2}}{yy^{\prime}}\left[f\left(x,x';y,y^{\prime};t,t'\right)\left(1+\frac{\left(\triangle x\right)^{2}}{2\beta_{H}^{2}}+\ldots\right)-g\left(x,x';y,y^{\prime};t,t'\right)\left(1+\frac{\left(\triangle t\right)^{2}}{2\beta_{H}^{2}}+\ldots\right)\right]\nonumber \\
 & = & \frac{1}{2yy^{\prime}}\left[f\left(x,x';y,y^{\prime};t,t'\right)\left(\triangle x\right)^{2}-g\left(x,x';y,y^{\prime};t,t'\right)\left(\triangle t\right)^{2}\right.\nonumber \\
 &  & \left.+2\beta_{H}^{2}\left(f\left(x,x';y,y^{\prime};t,t'\right)-g\left(x,x';y,y^{\prime};t,t'\right)\right)+\ldots\right]\nonumber \\
 & \rightarrow & \frac{\left(\triangle x\right)^{2}-\left(\triangle t\right)^{2}+y^{2}+y^{\prime2}}{2yy^{\prime}}.\label{eq:cond.3}
\end{eqnarray}

\noindent In contrast, when calculating the metric using
eqn. (\ref{eq:derive metric}), we observe that $f\left(x,x';y,y^{\prime};t,t'\right)$
and $g\left(x,x';y,y^{\prime};t,t'\right)$ enter $g_{xx}$ and $g_{tt}$.
However, we know that for large $\beta_{H}$, it must reduce to the asymptotic
AdS in the Poincare coordinates. Thus, we conclude that $f\left(x,x';y,y^{\prime};t,t'\right)$
and $g\left(x,x';y,y^{\prime};t,t'\right)$ are independent of $x,x'$
and $t,t'$. Note that $f$ and $g$ are dimensionless.  Therefore, we rewrite
the general expression of the geodesic length as

\vspace{1em}

\noindent\fbox{\begin{minipage}[t]{1\columnwidth - 2\fboxsep - 2\fboxrule}%
\begin{equation}
\cosh\left(\frac{L_{\mathrm{bulk}}}{R}\right)=\frac{\beta_{H}^{2}}{yy^{\prime}}\left[f\left(\frac{y}{\beta_{H}},\frac{y'}{\beta_{H}}\right)\cosh\left(\frac{\triangle x}{\beta_{H}}\right)-g\left(\frac{y}{\beta_{H}},\frac{y'}{\beta_{H}}\right)\cosh\left(\frac{\triangle t}{\beta_{H}}\right)\right],\label{eq:general expression of L}
\end{equation}
\end{minipage}}

\vspace{1em}

\noindent with

\begin{eqnarray}
f\left(\frac{y}{\beta_{H}},\frac{y^{\prime}}{\beta_{H}}\right) & = & 1+\mu_{1}\left(\frac{y}{\beta_{H}}+\frac{y^{\prime}}{\beta_{H}}\right)+\mu_{2}\left(\frac{y}{\beta_{H}}+\frac{y^{\prime}}{\beta_{H}}\right)^{2}+\ldots\nonumber \\
 &  & +\rho_{1}\left(\frac{yy^{\prime}}{\beta_{H}^{2}}\right)+\rho_{2}\left(\frac{yy^{\prime}}{\beta_{H}^{2}}\right)^{2}+\ldots,\nonumber \\
\nonumber \\
g\left(\frac{y}{\beta_{H}},\frac{y^{\prime}}{\beta_{H}}\right) & = & 1+\bar{\mu}_{1}\left(\frac{y}{\beta_{H}}+\frac{y^{\prime}}{\beta_{H}}\right)+\bar{\mu}_{2}\left(\frac{y}{\beta_{H}}+\frac{y^{\prime}}{\beta_{H}}\right)^{2}+\ldots\nonumber \\
 &  & +\bar{\rho}_{1}\left(\frac{yy^{\prime}}{\beta_{H}^{2}}\right)+\bar{\rho}_{2}\left(\frac{yy^{\prime}}{\beta_{H}^{2}}\right)^{2}+\ldots
\end{eqnarray}

\noindent Moreover, from eqn. (\ref{eq:cond.3}), matching the $y$ direction
of pure AdS$_{3}$ yields an important constraint:

\begin{equation}
f\left(\frac{y}{\beta_{H}},\frac{y^{\prime}}{\beta_{H}}\right)-g\left(\frac{y}{\beta_{H}},\frac{y^{\prime}}{\beta_{H}}\right)=\frac{1}{2\beta_{H}^{2}}\left(y^{2}+y^{\prime2}\right)+\mathcal{O}\left(\frac{1}{\beta_{H}^{4}}\right).
\end{equation}

\vspace{2em}

\noindent \textbf{Step 3:} When two endpoints of a geodesic coincide,
the geodesic length vanishes exactly. Substituting $x=x^{\prime}$, $y=y^{\prime}$,
and $t=t^{\prime}$ into eqn. (\ref{eq:general expression of L}),
we obtain
\begin{eqnarray}
\cosh\left(\frac{L_{\mathrm{bulk}}}{R}\right) & = & \frac{\beta_{H}^{2}}{y^{2}}\left[f\left(\frac{y}{\beta_{H}},\frac{y}{\beta_{H}}\right)-g\left(\frac{y}{\beta_{H}},\frac{y}{\beta_{H}}\right)\right]\nonumber \\
 & \rightarrow & 1,
\end{eqnarray}

\noindent which leads to

\begin{equation}
f\left(\frac{y}{\beta_{H}},\frac{y}{\beta_{H}}\right)-g\left(\frac{y}{\beta_{H}},\frac{y}{\beta_{H}}\right)=\frac{y^{2}}{\beta_{H}^{2}}.
\end{equation}

\vspace{2em}

\noindent \textbf{Step 4:}
In ref. \cite{Takayanagi:2011zk}, Takayanagi proposed a new holographic dual of the BCFT. It states that the phase transitions of EE relate to the topological change of the RT surface in the bulk. Based on this realization, the boundary of the BCFT will extend into the bulk and play a role in the end-of-the-world (ETW) brane \cite{Sully:2020pza}. The brane's tension corresponds to the boundary entropy of the BCFT, and the RT surface that is anchored between the ETW in the bulk and the BCFT on the boundary relates to the EE of the BCFT. The region enclosed by the ETW brane in the bulk and the BCFT on the boundary is asymptotically AdS. Therefore, the dual RT surfaces of the BCFT can be simply calculated between one point in the bulk and another point on the boundary in the AdS background without placing any new configuration, such as virtual branes that modify the bulk geometry \cite{Takayanagi:2011zk}. We will use this conclusion in this step because our method depends only on the RT surfaces and aims to fix the leading behaviors of bulk geometry.

From eqn. (\ref{eq:BCFT}), the BCFT provides
the EE of the half line. However we should replace $\Delta x$
by $\Delta x/2$ here because we use $\Delta x$ to represent the total
size of the entangled region,

\begin{equation}
S_{EE}=\frac{c}{6}\log\left(\frac{2\beta_{H}}{a}\sinh\frac{\triangle x}{2\beta_{H}}\right)\Longrightarrow L_{\mathrm{BCFT}}=R\log\left(\frac{2\beta_{H}}{a}\sinh\left(\frac{\triangle x}{2\beta_{H}}\right)\right).
\end{equation}
When $\triangle x\rightarrow\infty$, it becomes

\begin{equation}
L_{\mathrm{BCFT}}=R\log\left[\frac{\beta_{H}}{a}\exp\left(\frac{\triangle x}{2\beta_{H}}\right)\right].\label{eq:half 1}
\end{equation}

\begin{figure}[H]
\begin{centering}
\includegraphics[scale=0.4]{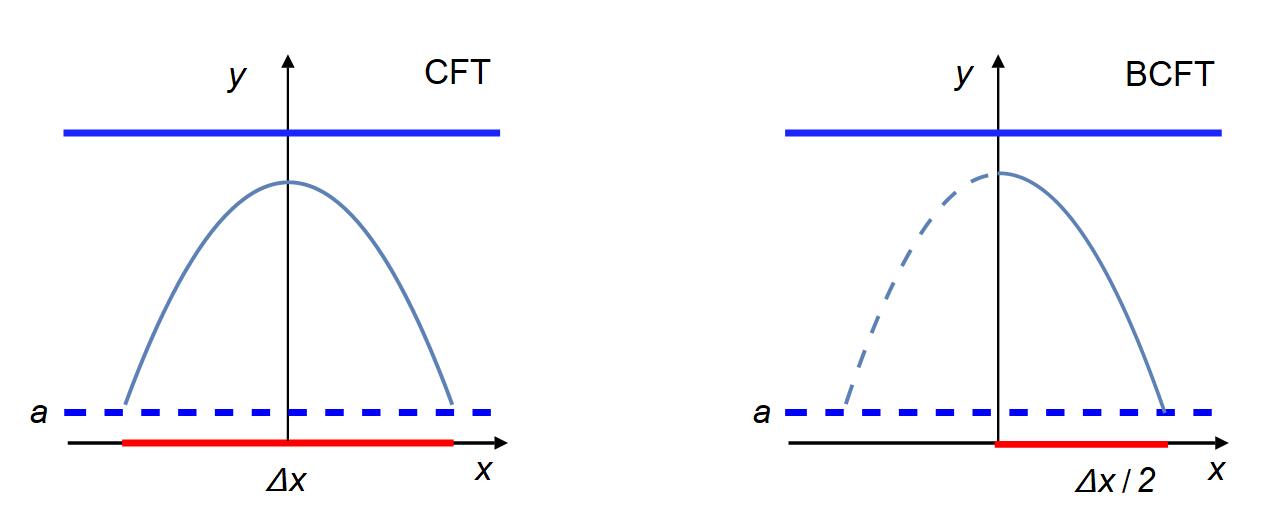}
\par\end{centering}
\centering{}\caption{The left-hand side image shows the geodesic identified from CFT
EE. The endpoints are fixed at $y=y^{\prime}=a$.
The solid curve in the right-hand side  image represents the geodesic
identified from the BCFT EE. The endpoints are fixed
at $y=a$ and $y'=\beta_{H}$, and the entangling length is $\Delta x/2$.
\label{fig:The-left-hand}}
\end{figure}

\noindent As shown in Fig. (\ref{fig:The-left-hand}), the geodesic
corresponding to this EE connects $y=a$ and $y'=\beta_{H}$.

\noindent In contrast, by using the general expression of the
geodesic length (\ref{eq:general expression of L}), we have two other
methods of  calculating the length of this geodesic. The first method is to
straightforwardly substitute $y=a$, $y'=\beta_{H}$, and $\Delta x/2\to\infty$
into (\ref{eq:general expression of L}) to obtain

\begin{equation}
L_{\mathrm{bulk}}\rightarrow L_{\mathrm{half1}}=R\log\left(\frac{\beta_{H}}{a}\,f\left(\frac{a}{\beta_{H}},\frac{\beta_{H}}{\beta_{H}}\right)\exp\left(\frac{\triangle x}{2\beta}\right)\right).\label{eq:half 2}
\end{equation}

\noindent We easily understand that this length is half
the geodesic length connecting $y=y^{\prime}=a$ and $\Delta x\to\infty$.
Therefore, the second method is

\begin{equation}
L_{\mathrm{bulk}}\rightarrow L_{\mathrm{half2}}=\frac{1}{2}\,R\log\left(\frac{\beta_{H}^{2}}{a^{2}}\,f\left(\frac{a}{\beta_{H}},\frac{a}{\beta_{H}}\right)\exp\left(\frac{\triangle x}{\beta_{H}}\right)\right).\label{eq:half 3}
\end{equation}

\begin{figure}[H]
\begin{centering}
\includegraphics[scale=0.4]{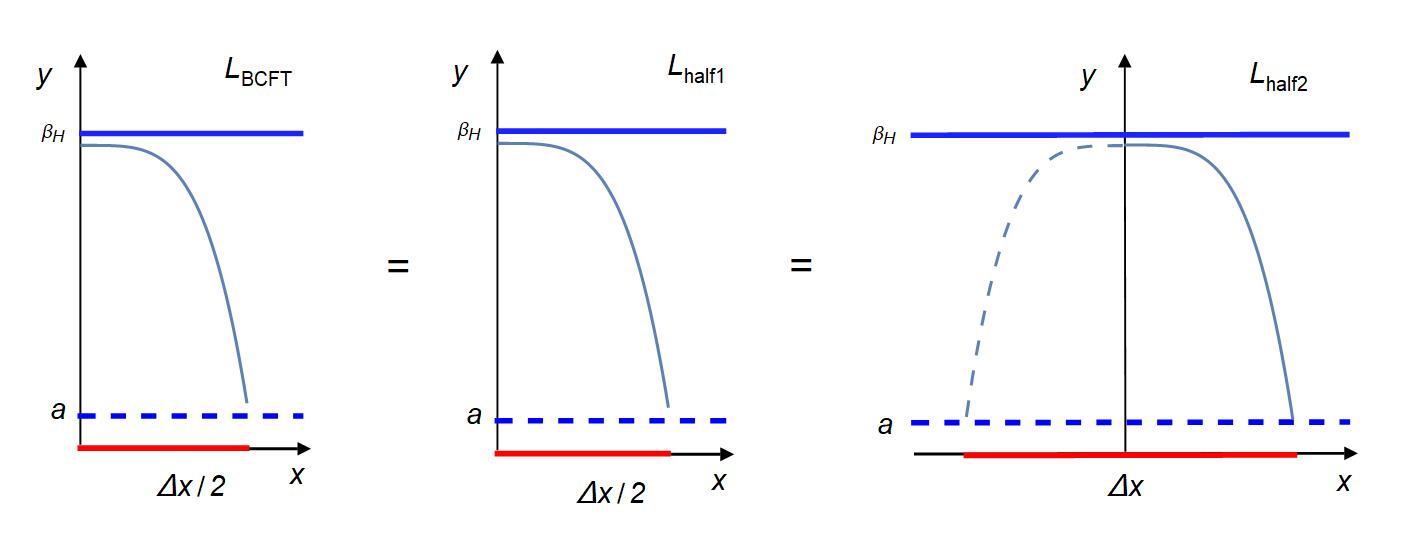}
\par\end{centering}
\centering{}\caption{The left-hand image is given by the BCFT. The middle image is obtained
from $L_{\mathrm{bulk}}$ by setting $y=a$, $y^{\prime}=\beta_{H}$
and $\triangle x/2\rightarrow\infty$. The right-hand image is also given
by $L_{\mathrm{bulk}}$ from a different perspective, by setting
$y=y^{\prime}=a$ and $\Delta x\to\infty$. The solid lines in all
the three pictures describe the same object. \label{fig:The-left-picture}}
\end{figure}

\noindent These three lengths in eqn. (\ref{eq:half 1}), (\ref{eq:half 2})
and (\ref{eq:half 3}) must be identical, as shown in Fig. (\ref{fig:The-left-picture}).
Thus, we obtain

\begin{equation}
f\left(\frac{a}{\beta_{H}},\frac{\beta_{H}}{\beta_{H}}\right)^{2}=f\left(\frac{a}{\beta_{H}},\frac{a}{\beta_{H}}\right)=1.
\end{equation}

\noindent The derivation of this constraint does not require $\beta_{H}\gg a$.
As long as $\beta_{H}$ is the upper bound of $y$, the derivation
is justified. Because $a$ is a varying cut-off not beyond $\beta_{H}$,
satisfying $0<a/\beta_{H}\le1,$ we can safely replace $\frac{a}{\beta_{H}}$
by $\frac{y}{\beta_{H}}$ to obtain:

\begin{equation}
f\left(\frac{y}{\beta_{H}},1\right)^{2}=f\left(\frac{y}{\beta_{H}},\frac{y}{\beta_{H}}\right)=1.
\end{equation}

\vspace{2em}

\noindent \textbf{Step 5:} An important lesson we learn from the
free CFT$_{2}$ case is that, to completely determine the
dual geometry, we must know the geodesic length between $a$ and
$\beta_{H}$ with $\triangle x=0$, i.e. the vertical geodesic. To
be consistent, this particular geodesic length must be provided by
the CFT$_{2}$ EE. In the free CFT$_{2}$, the IR
EE precisely fits the requirement. The finite
temperature CFT$_{2}$ does not have such IR EE.

Remarkably, we know that the finite
temperature CFT$_{2}$ and finite size CFT$_{2}$ have the same geometry ${\mathbb R}\times {\mathbf S}^1$. We can either interpret it as a CFT on a compact spatial interval of size $L_S$ or as a thermal CFT on the real line with the Euclidean time along the circle with the period  $\beta = L_S$, as explained in \cite{Calabrese:2004eu, Rangamani:2016dms} in detail. Therefore two CFTs are basically the same scenario and  have the same bulk dual. We are allowed to use the results from both CFTs to construct the dual bulk geometry with the identification
Therefore, we map the finite temperature system to a finite
size ($L_{S}$) system by replacing $\beta\rightarrow L_{S}$ and
impose the periodic boundary condition\footnote{To make the discussion simpler, we do not choose the equivalent replacement $L_S=i\beta_H$.}

\begin{equation}
\beta_{H}=\frac{\beta}{2\pi}\rightarrow\frac{L_{S}}{2\pi}.
\end{equation}

\noindent Therefore, the geodesic length between $a$ and $\beta_{H}=\frac{\beta}{2\pi}$
in the finite temperature system equals the one that connects $a$ and $\frac{L}{2\pi}$
in the finite size system:

\begin{eqnarray}
\mathrm{finite\;temperature} &  & \mathrm{finite\;size}\nonumber \\
L_{\mathrm{geodesic}}\left(a,\frac{\beta}{2\pi}\right) & = & L_{\mathrm{geodesic}}\left(a,\frac{L_{S}}{2\pi}\right).
\end{eqnarray}

\begin{figure}[H]
\begin{centering}
\includegraphics[scale=0.4]{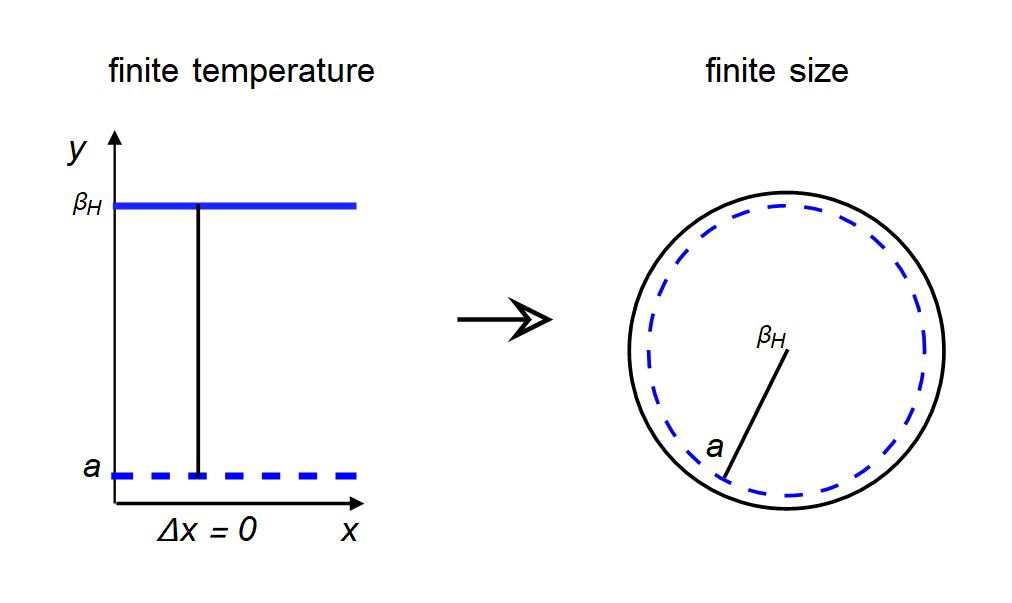}
\par\end{centering}
\centering{}\caption{\label{fig:The-geodesic-between}Geodesic between $y=a$ and $y^{\prime}=\beta_{H}$
at a finite temperature system mapped to a finite size system,
corresponding the radius of a circle from $y=a$ to $y=\frac{L}{2\pi}$.}
\end{figure}

\noindent Noting that $\frac{L_{S}}{2\pi}$ is the radius of the finite
size system with the circumference $L_{S}$, this geodesic
extends from boundary to the center of the circle, as shown in
Fig. (\ref{fig:The-geodesic-between}).

\noindent We know that the EE of a finite size system
is

\begin{equation}
S_{EE}=\frac{c}{3}\log\left(\frac{L_{S}}{\pi a}\sin\left(\frac{\pi\triangle x}{L_{S}}\right)\right),
\end{equation}

\noindent The maximal EE is achieved by splitting
the circle into two equal regions, $\Delta x=L_{S}/2$. The corresponding
geodesic is simply a diameter

\begin{equation}
S_{EE}=\frac{c}{3}\log\left(\frac{L_{S}}{\pi a}\right),\qquad L_{\mathrm{boundary}}=2R\log\left(\frac{L_{S}}{\pi a}\right).
\end{equation}

\noindent Thus, we can obtain the value we desire:

\begin{equation}
L_{\mathrm{radius}}=\frac{1}{2}L_{\mathrm{boundary}}=R\log\left(\frac{L_{S}}{\pi a}\right).
\end{equation}

\noindent We now map $L_{S}\to\beta$ to obtain the geodesic length between
$a$ and $\beta_{H}=\frac{\beta}{2\pi}$ in the finite temperature
system:

\begin{equation}
L=R\log\left(\frac{\beta}{\pi a}\right)=R\log\left(\frac{2\beta_{H}}{a}\right).
\end{equation}

\noindent Therefore, from the general expression (\ref{eq:general expression of L}),
as $x=x^{\prime}$, $t=t^{\prime}$, $y=a$, and $y^{\prime}=\beta_{H}$
, we obtain

\begin{eqnarray}
L_{\mathrm{boundary}} & = & R\log\left(\frac{2\beta_{H}}{a}\left[f\left(\frac{a}{\beta_{H}},\frac{\beta_{H}}{\beta_{H}}\right)-g\left(\frac{a}{\beta_{H}},\frac{\beta_{H}}{\beta_{H}}\right)\right]\right)\nonumber \\
 & \rightarrow & R\log\frac{2\beta_{H}}{a}.
\end{eqnarray}

\noindent Thus, we obtain

\begin{equation}
f\left(\frac{a}{\beta_{H}},\frac{\beta_{H}}{\beta_{H}}\right)-g\left(\frac{a}{\beta_{H}},\frac{\beta_{H}}{\beta_{H}}\right)=1.
\end{equation}

\vspace{2em}

\noindent For convenience, we summarize all the constraints we have
obtained for the general expression (\ref{eq:general expression of L})
of the geodesic length:

\begin{eqnarray}
 &  & f\left(\frac{y}{\beta_{H}},\frac{y^{\prime}}{\beta_{H}}\right)-g\left(\frac{y}{\beta_{H}},\frac{y^{\prime}}{\beta_{H}}\right)=\frac{1}{2\beta_{H}^{2}}\left(y^{2}+y^{\prime2}\right)+\mathcal{O}\left(\frac{y^{4}}{\beta_{H}^{4}}\right),\hspace{1em}\beta_{H}\gg y,y',\label{eq:R3}\\
\nonumber \\
 &  & f\left(\frac{y}{\beta_{H}},\frac{\beta_{H}}{\beta_{H}}\right)^{2}=f\left(\frac{y}{\beta_{H}},\frac{y}{\beta_{H}}\right)=1,\label{eq:R4}\\
 &  & f\left(\frac{y}{\beta_{H}},\frac{y}{\beta_{H}}\right)-g\left(\frac{y}{\beta_{H}},\frac{y}{\beta_{H}}\right)=\frac{y^{2}}{\beta_{H}^{2}},\label{eq:R5}\\
 &  & f\left(\frac{a}{\beta_{H}},\frac{\beta_{H}}{\beta_{H}}\right)-g\left(\frac{a}{\beta_{H}},\frac{\beta_{H}}{\beta_{H}}\right)=1.\label{eq:R6}
\end{eqnarray}

\noindent From eqn. (\ref{eq:R4}) and (\ref{eq:R6}), we obtain

\begin{equation}
g\left(\frac{a}{\beta_{H}},\frac{\beta_{H}}{\beta_{H}}\right)=0.
\end{equation}

\noindent Because $a$ is a varying quantity, $y$ or $y'=\beta_{H}$
must be a zero of $g(y/\beta_{H},y'/\beta_{H})$ . Moreover, $g$
must be symmetric for $y$ and $y'$. Therefore, the function form must be

\begin{equation}
g\left(\frac{y}{\beta_{H}},\frac{y'}{\beta_{H}}\right)\propto\left(1-\frac{y^{n}}{\beta_{H}^{n}}\right)^{\kappa}\left(1-\frac{y'^{n}}{\beta_{H}^{n}}\right)^{\kappa}\left(\cdots\right)
\end{equation}

\noindent Moreover, from eqn. (\ref{eq:R4}) and (\ref{eq:R5}),
we obtain

\begin{equation}
g\left(\frac{y}{\beta_{H}},\frac{y}{\beta_{H}}\right)=1-\frac{y^{2}}{\beta_{H}^{2}}
\end{equation}

\noindent Thus, we can easily to fix $n=2$ and $\kappa=1/2$ and

\begin{equation}
g\left(\frac{y}{\beta_{H}},\frac{y'}{\beta_{H}}\right)=\sqrt{\left(1-\left(\frac{y}{\beta_{H}}\right)^{2}\right)\left(1-\left(\frac{y^{\prime}}{\beta_{H}}\right)^{2}\right)}\left[1+\left(\frac{\Delta y}{\beta_{H}}\right)^{2}\left(\sigma_{1}+{\cal O}\left(\frac{y}{\beta}\right)\right)\right].\label{eq:g expansion}
\end{equation}

\noindent Similarly, from eqn. (\ref{eq:R4}), we obtain

\begin{equation}
f\left(\frac{y}{\beta_{H}},\frac{y'}{\beta_{H}}\right)=1+\left(\frac{\Delta y}{\beta_{H}}\right)^{2}\left(1-\frac{y^{m}}{\beta_{H}^{m}}\right)^{\delta}\left(1-\frac{y'^{m}}{\beta_{H}^{m}}\right)^{\delta}\left[\theta_{1}+{\cal O}\left(\frac{y}{\beta}\right)+\cdots\right],\label{eq:f expansion}
\end{equation}

\noindent where $m$, $\delta>0$ are some numbers.

\noindent The story is not over yet. To match eqn. (\ref{eq:R3}),
we must have $\sigma_{1}=\theta_{1}$. When applying eqn. (\ref{eq:derive metric})
to calculate the metric, noting that a limit $\Delta x,\Delta y,\Delta t\to0$
will be imposed after making the derivatives, we easily observe
that the terms proportional to $\left(\frac{\Delta y}{\beta_{H}}\right)^{2}$
contribute only to $g_{yy}$ but not to $g_{xx}$ and $g_{tt}$.
Therefore, based on eqn. (\ref{eq:g expansion}) and (\ref{eq:f expansion}),
without altering the derived metric, equivalently, we are free to
pack all the corrections into $g(y,y')$ and simply set $f(y,y')=1$.
Finally, we obtain

\begin{equation}
\cosh\left(\frac{L_{\mathrm{bulk}}}{R}\right)=\frac{\beta_{H}^{2}}{yy^{\prime}}\left[\cosh\left(\frac{\triangle x}{\beta_{H}}\right)-\sqrt{\left(1-\left(\frac{y}{\beta_{H}}\right)^{2}\right)\left(1-\left(\frac{y^{\prime}}{\beta_{H}}\right)^{2}\right)}\left(1+\left(\frac{\Delta y}{\beta_{H}}\right)^{2}\cdot{\cal O}\left(\frac{y}{\beta}\right)\right)\cosh\left(\frac{\triangle t}{\beta_{H}}\right)\right].
\end{equation}

\noindent Applying (\ref{eq:derive metric}), we obtain the metric
of the BTZ black hole:

\begin{equation}
ds^{2}=\frac{R_{AdS}^{2}}{y^{2}}\left(-\left(1-\frac{y^{2}}{\beta_{H}^{2}}\right)dt^{2}+dx^{2}+\left(1-\frac{y^{2}}{\beta_{H}^{2}}\right)^{-1}\left(1+{\cal O}\left(\frac{y}{\beta_{H}}\right)\right)dy^{2}\right).
\end{equation}
Because the finite size system has the same topology as the finite temperature
one, a simple method of fixing the leading behavior of the dual geometry of the finite size system
is to use the transformation $\beta=-iL$, which leads to the pure
AdS$_{3}$ in the global coordinate. Similarly, we can fix the leading behaviors of
the dual geometries of CFT$_{2}$ with topological defects under transformation
$\beta=-iL/\gamma_{\mathrm{con}}$.

\section{Discussions and conclusion}

In this final section, we discuss our results and provide several inspirations
as well as conjectures. In summary, we have demonstrated an approach to
fixing the leading behaviors of three dimensional dual geometries, such as asymptotic AdS$_{3}$
and BTZ black hole, from the EEs of $\mathrm{CFT}_{2}$.
Our derivation relies only on the holographic principle without any
assumptions about AdS/CFT and bulk geometry. The steps of the method are as
follows:
\begin{enumerate}
\item Identify the energy cut-off as an extra dimension.
\item Identify the EE with the geodesic length of the unknown
dual geometry. The geodesics are attached on the boundary.
\item Write down the bulk geodesic length by making the most general extension
of the geodesic ending on boundary to include the extra dimension.
\item Use properties a geodesic must respect, say, zero length for coincide
endpoints, to impose constraints on the bulk geodesic function form.
\item Use the IR-like EE representing a geodesic whose
one endpoint stands on the boundary and another stretches into the
bulk of the unknown geometry, to restrict the bulk geodesic function
form.
\item Apply eqn. (\ref{eq:derive metric}) to fix the leading behavior of the metric.
\end{enumerate}
Our approach, from the derivation of the BTZ
black hole, may apply for all three dimensional geometries. As we have explained,
because they have the same topology as the finite temperature CFT$_{2}$,
the finite size CFT$_{2}$ and CFT$_{2}$ with topological defects
can be easily determined using simple transformations, although an independent
parallel derivation is desirable. Basically, we need only to consider
one representative for each topology. Therefore, the next non-trivial steps
involve investigating the chiral CFT$_{2}$ or finite size thermal CFT$_{2}$.
Probably, the only obstacle is to determine the IR-like EEs
from the CFT side. The existence of the IR-like EEs
is unquestionable, but it might be difficult to calculate from the
CFT side. A compromise is to borrow the IR-like EEs
from holographic calculation, if not so strict. For CFT living on
surfaces beyond the torus, we believe the derivation can still performed, but we must know the EEs, both UV and IR-like,
which are difficult  to obtain from the CFT side. In addition, it would
be of significant interest to consider the CFT$_{2}$ which are dual to
sourced gravity. We may learn more non-trivial things from these examples.

Our derivation shows that the bulk geometry cannot be determined using UV EE only. For a higher dimensional
case $d+1>3$, the scenario is complicated. One reason is that
no method is available to calculate the metric from minimal surfaces.
Another challenge is that even for a single topology, many
inequivalent gravitational structures in higher dimensions. We are
not certain if other subtleties ill occur.

An interesting question is, \emph{Is it necessary to identify
the EE with the geodesic to fix the leading behavior of the metric?}
The answer appears to be negative. As we know, the arguments of the EE
 include both spacetime directions $t$, $x$ as well as energy
cut-offs such as $a$, $\xi$, $\beta$....After identifying the energy
scale as an extra dimension $y$, we can introduce a generalized
EE $\mathcal{S}_{EE}\left(t,t';x,x';y,y'\right)$
as follows:
\begin{itemize}
\item We denote the energy generated dimension as $y$. Therefore, the energy cut-offs
are different values on the dimension $y$.
\item Because $y$ is on the same footing as the ordinary spacetime directions
$x$, $t$, it is natural to generalize $S_{EE}(t,t';x,x';a,\beta,\cdots)$
to $\mathcal{S}_{EE}\left(t,t';x,x';y,y'\right)$ in the most generic
manner.\\$\Longrightarrow$This step corresponds to extending the boundary
attached geodesic to the bulk geodesic.
\item Under various limits, the generalized EE $\mathcal{S}_{EE}\left(t,t';x,x';y,y'\right)$
should reproduce all the EEs of a specified
CFT, such as the UV or IR-like EEs . \\$\Longrightarrow$This
step corresponds to using various EEs to determine
the behaviors of the regular functions in our approach.
\item The generalized EE $\mathcal{S}_{EE}\left(t,t';x,x';y,y'\right)$
should be renormalized because the infinities of QFT are caused by
energy, which is now a new dimension. \\$\Longrightarrow$ This step
corresponds to demanding the vanished geodesic length for coincident endpoints
and other consistencies.
\end{itemize}
Therefore, our previous calculations naturally
fit the procedure. Now, we immediately have an interesting equation:

\begin{equation}
\frac{1}{2}\mathcal{S}_{EE}^{2}\left(x;x+dx\right)=g_{ij}\left(x\right)dx^{i}dx^{j}+\mathcal{O}\left(dx^{2}\right).\label{eq:new interpretation}
\end{equation}

\noindent All the derivations in this paper can be expressed in this pattern.
Consequently all GR quantities, such as the connection, Riemann tensor
etc. can be subsequently constructed. This equation indicates some
new interpretations:
\begin{itemize}
\item Spacetime is not an emergent structure from quantum entanglement,
but is quantum entanglement itself, simply viewed from a different
angle.
\item Quantum entanglement with different lengths knits the spacetime.
\end{itemize}

\noindent
Moreover, we wish to clarify two significant reasons for our derivations appearing  heavy:

\begin{enumerate}
\item The advantages of our method is to also cover the time-like
direction of the spacetime metric naturally. This is very difficult
because we know only the information of the lower dimensional theories.

\item Our objective is to determine not only the linear order but also the singularity and event horizon of BTZ spacetime. The behaviors of black hole's singularity and event horizon cannot be extracted using the leading term of the spacetime metric directly. This is why we use more results of EEs, and aim to fix more accurate leading behaviors of bulk geometries.
\end{enumerate}

\noindent
Therefore, in this paper, our aim is not to explain how bulk
geometry emerges from EEs, or to derive the bulk
dynamics (Einstein's equation) from the boundary theory but only to show that
the EEs of $\mathrm{CFT}_{2}$ are sufficient to fix
the leading behaviors of the bulk spacetime geometries.

Another point deserves a special emphasis. Our results demonstrate
that when we treat the energy scale as a usual space-like dimension,
the CFT contains almost all the classical information of the dual
geometry, at least for $d=2$. In AdS/CFT correspondence, to compare
the correlation functions of the dual theories, we take limits to
push the AdS$_{d+1}$ bulk-to-bulk correlation function onto the boundary
and then match the CFT$_{d}$ correlation function \cite{Witten:1998qj}.
However, the method of lifting the CFT$_{d}$ correlation function into the bulk
directly is still an open question. Our derivations show that
when we treat the energy scale as an extra dimension, after imposing
some consistent constraints, the bulk-to-bulk
correlation function from the boundary-to-boundary can be derived. Thinking it over,
we observe that two equations govern the dynamics of operators in QFT:
the Callan-Symanzik (RG) equation and equation of motion (EOM).
The Callan-Symanzik equation informs us how the operators evolve with
respect to energy scales. The EOM determines the evolution of the
operators with respect to spacetime coordinates. Therefore, logically,
we can naturally conjecture that

\vspace{2em}

\noindent\fbox{\begin{minipage}[t]{1\columnwidth - 2\fboxsep - 2\fboxrule}%
\[
\mathrm{Callan-Symanzik\;(RG)\;equation\qquad+\qquad EOM\;on\;flat\qquad=\qquad EOM\;in\;the\;bulk},
\]
\end{minipage}}

\vspace{2em}

which implies a unification of the RG equation and field EOM.

\vspace{5mm}

\noindent {\bf Acknowledgements}
We are deeply indebted to Bo Ning  for many illuminating discussions and suggestions.  We are also very grateful to Q. Gan, S. Kim, J. Lu,  H. Nakajima, S. Ying, and S. He for very helpful discussions and suggestions. This work is supported in part by the NSFC (Grant No. 12105191, 12275183 and 12275184). 


\begin{thebibliography}{99}

\bibitem{Maldacena:1997re}
J.~M.~Maldacena,   ``The Large N limit of superconformal field theories and supergravity,''   Int.\ J.\ Theor.\ Phys.\  {\bf 38}, 1113 (1999)   [Adv.\ Theor.\ Math.\ Phys.\  {\bf 2}, 231 (1998)]   doi:10.1023/A:1026654312961   [hep-th/9711200].   




\bibitem{Ryu:2006bv}
S.~Ryu and T.~Takayanagi,   ``Holographic derivation of entanglement entropy from AdS/CFT,''   Phys.\ Rev.\ Lett.\  {\bf 96}, 181602 (2006)   doi:10.1103/PhysRevLett.96.181602   [hep-th/0603001].   

\bibitem{Ryu:2006ef}    S.~Ryu and T.~Takayanagi,   ``Aspects of Holographic Entanglement Entropy,''   JHEP {\bf 0608}, 045 (2006)   doi:10.1088/1126-6708/2006/08/045   [hep-th/0605073].   

\bibitem{Hubeny:2007xt}    V.~E.~Hubeny, M.~Rangamani and T.~Takayanagi,   ``A Covariant holographic entanglement entropy proposal,''  JHEP {\bf 0707}, 062 (2007)  doi:10.1088/1126-6708/2007/07/062 [arXiv:0705.0016 [hep-th]]. 

\bibitem{Rangamani:2016dms}    M.~Rangamani and T.~Takayanagi,   ``Holographic Entanglement Entropy,''   arXiv:1609.01287 [hep-th].   






\bibitem{VanRaamsdonk:2009ar}    M.~Van Raamsdonk,   ``Comments on quantum gravity and entanglement,''   arXiv:0907.2939 [hep-th].   

\bibitem{VanRaamsdonk:2010pw}    M.~Van Raamsdonk,   ``Building up spacetime with quantum entanglement,''   Gen.\ Rel.\ Grav.\  {\bf 42}, 2323 (2010)   [Int.\ J.\ Mod.\ Phys.\ D {\bf 19}, 2429 (2010)]   doi:10.1007/s10714-010-1034-0, 10.1142/S0218271810018529   [arXiv:1005.3035 [hep-th]].   


\bibitem{Maldacena:2013xja}    J.~Maldacena and L.~Susskind,   ``Cool horizons for entangled black holes,''   Fortsch.\ Phys.\  {\bf 61}, 781 (2013)   doi:10.1002/prop.201300020   [arXiv:1306.0533 [hep-th]].   

\bibitem{Swingle:2009bg}    B.~Swingle,   ``Entanglement Renormalization and Holography,''   Phys.\ Rev.\ D {\bf 86}, 065007 (2012)   doi:10.1103/PhysRevD.86.065007   [arXiv:0905.1317 [cond-mat.str-el]].   

\bibitem{Czech:2015qta}
B.~Czech, L.~Lamprou, S.~McCandlish and J.~Sully,   ``Integral Geometry and Holography,''   JHEP {\bf 1510}, 175 (2015)   doi:10.1007/JHEP10(2015)175   [arXiv:1505.05515 [hep-th]].   



\bibitem{Wang:2017bym}    P.~Wang, H.~Wu and H.~Yang,   ``AdS$_3$ metric from UV/IR entanglement entropies of CFT$_2$,''   arXiv:1710.08448 [hep-th].   

\bibitem{Poisson:2011nh}    E.~Poisson, A.~Pound and I.~Vega,   ``The Motion of point particles in curved spacetime,''  Living Rev.\ Rel.\  {\bf 14}, 7 (2011) doi:10.12942/lrr-2011-7  [arXiv:1102.0529 [gr-qc]].  



\bibitem{Calabrese:2004eu}    P.~Calabrese and J.~L.~Cardy,   ``Entanglement entropy and quantum field theory,''   J.\ Stat.\ Mech.\  {\bf 0406}, P06002 (2004)   doi:10.1088/1742-5468/2004/06/P06002   [hep-th/0405152].   

\bibitem{Calabrese:2009qy}    P.~Calabrese and J.~Cardy,   ``Entanglement entropy and conformal field theory,''   J.\ Phys.\ A {\bf 42}, 504005 (2009)   doi:10.1088/1751-8113/42/50/504005   [arXiv:0905.4013 [cond-mat.stat-mech]].   

\bibitem{Takayanagi:2011zk} T.~Takayanagi, ``Holographic Dual of BCFT,'' Phys. Rev. Lett. \textbf{107}, 101602 (2011) doi:10.1103/PhysRevLett.107.101602 [arXiv:1105.5165 [hep-th]]. 

\bibitem{Sully:2020pza} J.~Sully, M.~Van Raamsdonk and D.~Wakeham, ``BCFT entanglement entropy at large central charge and the black hole interior,'' [arXiv:2004.13088 [hep-th]]. 





\bibitem{Witten:1998qj}
E.~Witten,   ``Anti-de Sitter space and holography,''   Adv.\ Theor.\ Math.\ Phys.\  {\bf 2}, 253 (1998)   doi:10.4310/ATMP.1998.v2.n2.a2   [hep-th/9802150].   

\end{thebibliography}
\end{document}